\documentclass[%
 reprint,
superscriptaddress,
 amsmath,amssymb,
 aps,
floatfix,
]{revtex4-2}

\usepackage{graphicx}
\usepackage{dcolumn}
\usepackage{bm}

\usepackage[colorlinks,
           urlcolor=blue,
           linkcolor=blue,
           anchorcolor=blue,
           citecolor=blue
           ]{hyperref}
\usepackage{float}
\usepackage{xcolor}
\usepackage{soul}

\begin{document}

\preprint{APS/123-QED}


\title{Temperature-Dependent Full Spectrum Optical Responses of \\ Semiconductors from First Principles}


\author{Zherui Han}
 \affiliation{School of Mechanical Engineering and the Birck Nanotechnology Center,\\
Purdue University, West Lafayette, Indiana 47907-2088, USA}
\author{Changkyun Lee}
 \affiliation{School of Electrical and Computer Engineering and the Birck Nanotechnology Center,\\
Purdue University, West Lafayette, Indiana 47907-2088, USA}
\author{Jiawei Song}
 \affiliation{School of Materials Engineering,
Purdue University, West Lafayette, Indiana 47907-2088, USA}
\author{Haiyan Wang}
 \affiliation{School of Materials Engineering,
Purdue University, West Lafayette, Indiana 47907-2088, USA}
\author{Peter Bermel}
 \affiliation{School of Electrical and Computer Engineering and the Birck Nanotechnology Center,\\
Purdue University, West Lafayette, Indiana 47907-2088, USA}
\author{Xiulin Ruan}%
 \email{ruan@purdue.edu}
 \affiliation{School of Mechanical Engineering and the Birck Nanotechnology Center,\\
Purdue University, West Lafayette, Indiana 47907-2088, USA}

\date{\today}

\begin{abstract}
From ultraviolet to mid-infrared region, light-matter interaction mechanisms in semiconductors progressively shift from electronic transitions to phononic resonances and are affected by temperature. Here, we present a parallel temperature-dependent treatment of both electrons and phonons entirely from first principles, enabling the prediction of full-spectrum optical responses. At elevated temperatures, \textit{ab initio} molecular dynamics is employed to find thermal perturbations to electronic structures and construct effective force constants describing potential landscape. Four-phonon scattering and phonon renormalization are included in an integrated manner in this approach. As a prototype ceramic material, cerium dioxide (CeO$_2$) is considered in this work. Our first-principles calculated refractive index of CeO$_2$ agrees well with measured data from literature and our own temperature-dependent ellipsometer experiment.

\end{abstract}

\maketitle


Study into temperature-dependent light-matter interactions is generally missing from theoretical side, as most optical applications only concern room temperature and constant optical parameters are always assumed. Following a surge in recent research into high-temperature or transient nanophotonics, plasmonics and near-field radiation, an understanding of temperature evolution of optical properties is now generally needed~\cite{ThermalRec2010PRL,GoldOptical2016,TNoptical2017}. Beside the interest of uncovering the physical mechanism behind this temperature evolution, accurate prediction also has technological importance: in thermophotovaltaics (TPV)~\cite{TPV2022}, heat-assisted magnetic recording technique (HAMR)~\cite{HAMR2014}, and thermal barrier coatings (TBC)~\cite{TBC2002}, high temperature optical properties are crucial for their realizations and performances.

Optical properties of metals at elevated temperatures have been studied from both experiments and theories~\cite{Silver2002,GoldOptical2016,TplasmonicPRB2016,SilverOptical2017}. Challenges are greater for semiconductors modeling since various elementary excitations or quasiparticles and their interplay need to be captured as a function of temperature~\cite{SemiOptics,RadiativePhonon2022}. A even greater challenge is to have full spectrum predictions with temperature dependence as various technologies do concern both a wide range of photon wavelengths and working temperatures, and the mechanism shifts from \textit{electron-mediated} to \textit{phonon-mediated} depending on photon wavelength. Generally, the optical properties depend on the electromagnetic field coupling with various model oscillators in solids. In semiconductors these oscillators can be excitons, optical phonons and potentially plasmons, with different coupling strength between the oscillator and electromagnetic field, and different oscillating frequency and damping strength of the resonance~\cite{SemiOptics}. In ultraviolet, visible and near infrared (UV-Vis-NIR) range, the primary interaction is between the electrons and photon (e.g., interband electronic excitations), while in the mid-infrared (MIR) region where incident photon energy is much lower, IR-active phonon resonances play the major role. Earlier studies~\cite{dieGeprb1984,epsilonPRB2008,T-excitonprl2008,eprenormprl2010} limited to wavelengths smaller than $1.2~\mathrm{\mu m}$ have elucidated the dynamical effect in \textit{electron-mediated} optical responses. Recent work~\cite{RadiativePhonon2022} in weak-coupling regime of the two-dimensional (2D) systems finds similar research gap in radiative linewidths, where 2D excitons are extensively studied but not phonons. The fact that a complete and rigorous first-principles approach is still generally lacking motivates us to investigate both \textit{electron-mediated} and \textit{phonon-mediated} optical processes as a function of temperature in general semiconductors.

This work considers the recent advances in first-principles techniques and temperature-dependent theories to enable a parallel, broadband treatment of both photon-electron and photon-phonon interactions in Cerium dioxide (CeO$_2$), a dielectric ceramic. The intrinsic resemblance of basic oscillators in optical responses make it possible for us to conceive a general and parallel approach. In particular, we capture the thermal perturbations by performing \textit{ab initio} molecular dynamics (AIMD) simulations to represent the statistical variations of physical quantities at a certain temperature. We then obtain temperature-dependent electronic and phononic structures by averaging the snapshots in AIMD or fitting an effective potential to get their renormalized energies, respectively. Dielectric function can further be calculated from electronic transitions~\cite{opticsVASP2006} in UV-Vis-NIR and application of Lorentz oscillator model~\cite{zhang2020nano} in MIR region. The key physical parameters involved in this process are band structure and phonon self-energy (both real and imaginary part), and we are able to resolve their temperature dependence concurrently. Specifically, we include phonon renormalization and four-phonon scattering~\cite{han2022prl} in the calculation of phonon energy and damping factor. In this study, we choose crystal CeO$_2$ as a benchmark material considering its broad technological importance, especially in high temperature applications such as oxide fuel cells, gas sensors~\cite{SOFC2011,CeriaThinFilm2020}, but the methodology above can be extended to dielectrics and semiconductors in general. We find that the imaginary part of the dielectric function increases, and refraction peak undergoes a red shift and a reduction in peak value with rising temperatures. Due to the reduction of the band gap, lower energy photons can be absorbed as the temperature increases. For wavelength longer than 500~nm, refractive index increases with temperature. Our own temperature-dependent ellipsometer measurements confirm the temperature trend of our calculated refractive index and validates our theoretical approach. This paper is organized as follows. After a brief discussions on computational techniques, we describe the process of obtaining the electronic band structure and phonon dispersions as a function of temperature. Then we present the computation of dielectric functions that span from UV-Vis-NIR to MIR range as a function of temperature. Finally we compare our temperature-dependent full spectrum refractive index with our own measurements.

The dielectric function, representing the optical responses of system, is computed by perturbation theory and Lorentz oscillator model for UV-Vis-NIR and MIR region, respectively. The expression for the imaginary part of dielectric tensor $\rm Im[\epsilon(\omega)]_{\alpha \beta}$ due to electronic transitions is~\cite{opticsVASP2006}
\begin{equation}
    \begin{split}
    \rm Im[\epsilon(\omega)]_{\alpha \beta}=\frac{4 \pi^{2} e^{2}}{\Omega} \lim _{q \rightarrow 0} \frac{1}{q^{2}} \sum_{c, v, \mathbf{k}} 2 w_{\mathbf{k}} \delta\left(\epsilon_{c \mathbf{k}}-\epsilon_{v \mathbf{k}}-\omega\right) \times \\ \left\langle u_{c \mathbf{k}+\mathbf{e}_{\alpha} q} \mid u_{v \mathbf{k}}\right\rangle\left\langle u_{v \mathbf{k}} \mid u_{c \mathbf{k}+\mathbf{e}_{\beta} q}\right\rangle,
    \end{split}
    \label{UV-optical}
    \end{equation}
where $\mathbf{e}$ denotes the unit vectors for three Cartesian directions $\alpha\beta\gamma$, $\Omega$ is the unit cell volume, $w_{\mathbf{k}}$ is the weights for $k-$point $\mathbf{k}$, $\epsilon$ is the electron energy with $c$ and $v$ denote conduction band and valence band, respectively. $u_{c \mathbf{k}}$ is the cell periodic part of the orbitals at $\mathbf{k}$. The real part of the dielectric function $\rm Re[\epsilon(\omega)]$ is further calculated by the Kramers-Kronig transformation. In CeO$_2$, $\epsilon(\omega)$ is isotropic. The above expression requires band energies at finite temperature. Turning to MIR region with phonon resonances, $\epsilon(\omega)$ is determined by a four-parameter Lorentz oscillator model~\cite{Barker1984,fourparameter2001}:
\begin{equation}
    \epsilon(\omega)=\epsilon_{\infty}\prod_{m} \frac{\omega_{m, \mathrm{LO}}^{2}-\omega^{2}+i\omega \gamma_{m, \mathrm{LO}}}{\omega_{m, \mathrm{TO}}^{2}-\omega^{2}+i\omega \gamma_{m, \mathrm{TO}} },
    \label{IR-optical}
\end{equation}
where $\epsilon_{\infty}$ is the dielectric constant at high-frequency limit that can be calculated by perturbation theory. LO and TO in the subscript denote the longitudinal and transverse optical phonons, respectively. $\omega_{m}$ is the resonance phonon frequency and $\gamma_{m}$ is the phonon damping factor corresponding to the $m-$th IR-active phonon modes, and $\epsilon(\omega)$ is the summation of all IR-active phonon resonances. This damping factor $\gamma$ is related to the phonon-phonon scattering rate $\tau^{-1}$ by $\gamma=(\tau^{-1}/2\pi)$. Such model requires temperature-dependent IR-active phonon energies and their scattering rates. To summarize, to compute dielectric function for the full spectrum, one needs temperature-dependent energy carrier energies and their coupling with photons. In this study, we compute these quantities from first principles.

All first-principles calculations are performed in in the framework of Density Functional Theory (DFT) or Density Functional Perturbation Theory (DFPT) as implemented in Vienna \textit{Ab initio} Simulation Package (VASP)~\cite{VASP1993}, including AIMD simulations and single-point calculation using finite difference method to obtain anharmonic interatomic force constants. Phononic structure and phonon damping factor are calculated utilizing \texttt{Phonopy}~\cite{phonopy} and \texttt{FourPhonon} package~\cite{shengbte,han2021fourphonon}. A known drawback of standard DFT for ceria-based materials is the poor description of Ce $4f$ state, and this is normally remedied by adding a Hubbard $U$ term~\cite{HubbardU1991}. We follow the suggestion in Ref.~\cite{LDA+UPRB2007} and use the local density approximation with Hubbard parameter (LDA+U) approach by setting $U_{\rm eff}=10~\rm eV$ in this study to correct our band structure. The simulated band gap (2.97~eV) is within the reported experimental range~\cite{wang2013ceo2bandgap,2016ceo2bandgap}. We note that another approach to study the quasiparticle properties is the \textit{ab initio} GW method~\cite{GW1986}, where the electron self-energy is linearly expanded in the screened Coulomb interaction ($W$). Coupled with the Bethe–Salpeter equation (BSE)~\cite{BSE1998,BSE2000}, this approach is proven to be successful to evaluate band gaps and optical properties in various material systems~\cite{BerkeleyGW}. However, previous study~\cite{GWinCeO2} suggests that such GW+BSE approach without Hubbard-like term does not well reproduce the band gap of CeO$_2$. In this work we use LDA+U method instead. Further computational details are presented in Supplemental Materials~\cite{supply}.

\begin{figure*}[ht]
    \centering
    \includegraphics[width=6.8in]{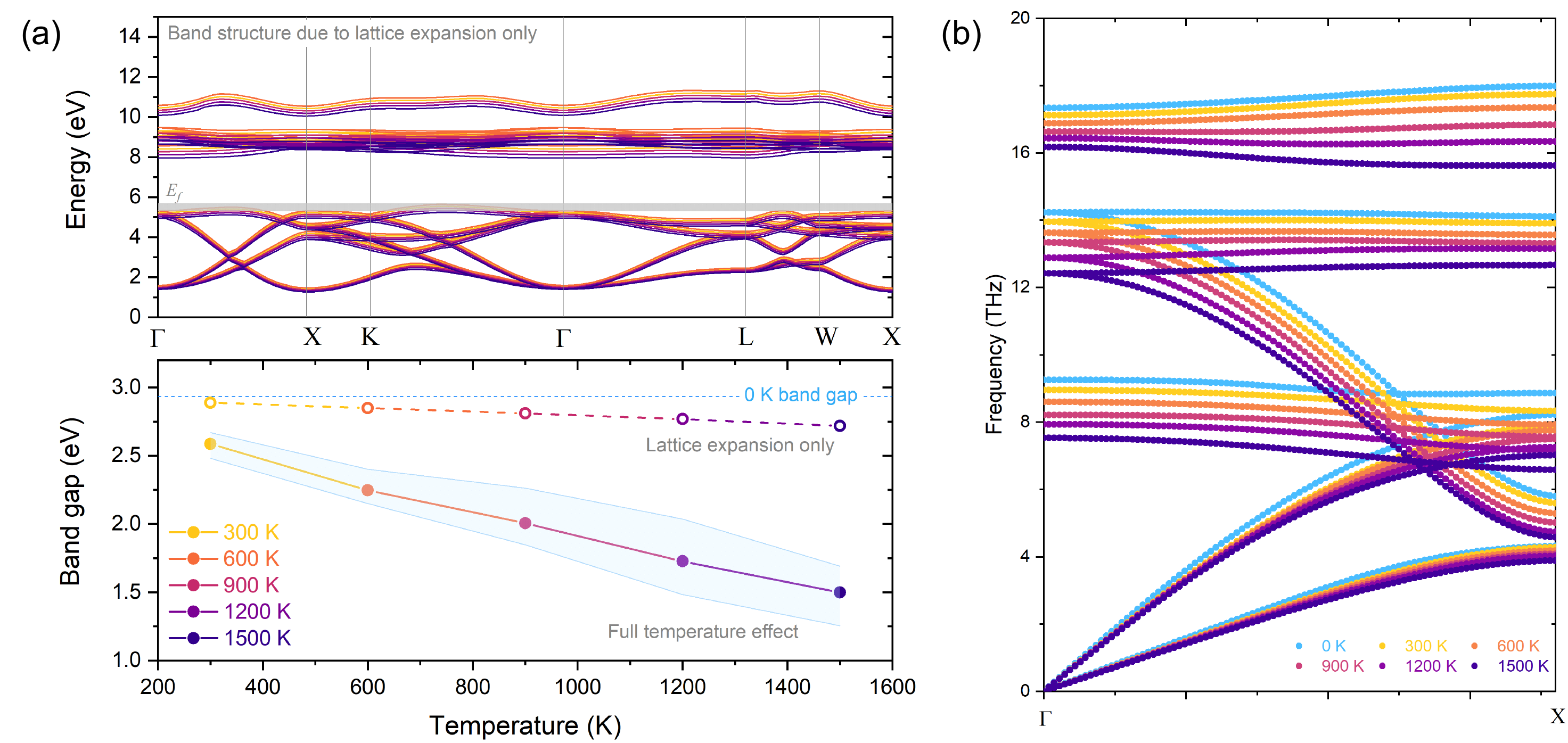}
    \caption{Temperature-dependent band structure and phonon dispersions. (a) Band structure with lattice expansion only (top panel), and the band gap as a function of temperature (bottom panel) when only lattice expansion is considered (dashed line) or dynamical effect is included (solid line). In top panel, the band structure at each $T$ is obtained by non-self-consistent DFT calculations at relaxed unit cell structure that comes from AIMD simulations under NPT ensemble. In the bottom panel, data points of full effect are the averaged band gap values calculated from 10 snapshots in AIMD simulations and the shaded area is the span of those 10 gap energies at each $T$. DFT band gap at 0~K is shown in blue dotted vertical line as a reference. (b) Phonon dispersions at finite $T$. All phonon energies are reduced with rising temperatures. Longitudinal and transverse optical phonon splitting is treated by considering long-range interactions using Born effective charge calculated by DFPT~\cite{phonopy}.}
    \label{dispersion}
\end{figure*}

A central issue is how we incorporate temperature into dispersion relations of two different energy carriers and obtain renormalized energy. Temperature renormalization of electronic band structure originates from thermal expansion and electron-phonon interactions (EPI)~\cite{Allen1981Bandgap}. One theoretical approach~\cite{PRB2019PbTe} to renormalize the band structure is to treat these two effects analytically by computing thermally expanded lattice and evoking Allen-Heine-Cardona (AHC) formalism~\cite{Allen1981Bandgap,ponce2015band}, respectively. This analytical approach conveys a clear physical picture but is limited to harmonic effects~\cite{epi2017revmodphys}. 
To have a consistent methodology for both electron and phonon renormalization, we need to capture anharmonic effect. AIMD simulation naturally considers thermal disorder~\cite{KimPRBband} and in the following discussion we would detail our parallel approach to renormalize both electron and phonon self-energies using AIMD. We start by bringing the system to a certain temperature $T$ (up to 1500~K) using AIMD under a NPT ensemble with zero external pressure. The simulation is performed on a supercell structure consisting of 192 atoms constructed by $4\times4\times4$ primitive cells. After reaching equilibrium, we find the relaxed structure at $T$. For CeO$_2$, we simulate the linear thermal expansion coefficient to be  $\alpha=1.05\times10^{-5}~\mathrm{K}^{-1}$ while the experimental reported value is around $1.16\times10^{-5}~\mathrm{K}^{-1}$~\cite{schwabExpTEC}. On top of this relaxed structure at every $T$, we perform AIMD under a NVT ensemble to sample the thermal perturbations. For electronic structure, we randomly choose 10 snapshots of NVT simulations and we calculate their band energies individually. This part of the treatment is similar to Ref.~\cite{KimPRBband,bandgapAPL} as we find $T$-dependent band gap by averaging snapshots, but the supercell system in our calculation is much larger. Figure~\ref{dispersion}(a) shows our calculated band structure and band gap due to thermal expansion only and the band gap due to temperature disorder altogether. The averaged band gap decreases with temperature, which is expected in semiconductors. We observe from the band structure with relaxed atomic structure at certain $T$ (see top panel and dashed line in the bottom panel of Fig.~\ref{dispersion}(a)) that the reduction due to lattice expansion is quite small, suggesting that in CeO$_2$ the dynamical effect from EPI is more pronounced than the lattice expansion. Also, the contribution from EPI is stronger at higher temperatures. Our calculated band gap evolution has a linear slope of $-9.53\times10^{-4}~\rm eV/K$, in agreement with a recent measurement up to 800~K~\cite{bandgapJCP2020} which is reported to be $-9.76\times10^{-4}~\rm eV/K$. This agreement of temperature evolution supports our AIMD approach to capture the dynamical effect of electronic structure. Now we turn to phonon dispersions that is relevant to resonances in MIR range. For phonons, we apply a Temperature Dependent Effective Potential method (TDEP) that uses all steps in NVT simulations to construct an effective harmonic force constants (HFCs)~\cite{TDEP2013} that can best describe the potential landscape at certain $T$. This effective HFCs then intrinsically includes the effect of anharmonic phonon-phonon interactions on the phonon frequency~\cite{TDEP2018PNAS} and renormalize the phonon energies. Note that the structure at certain $T$ is the relaxed structure. Fig.~\ref{dispersion}(b) presents our calculated phonon dispersions at a function of temperature using the effective HFCs. High temperature softens the phonon energies, especially the optical phonon branches. Since IR optical responses reply on IR-active phonon modes, this softening is expected to shift resonance peak observed in optical spectrum.

\begin{figure}[h]
    \centering
    \includegraphics[width=3.2in]{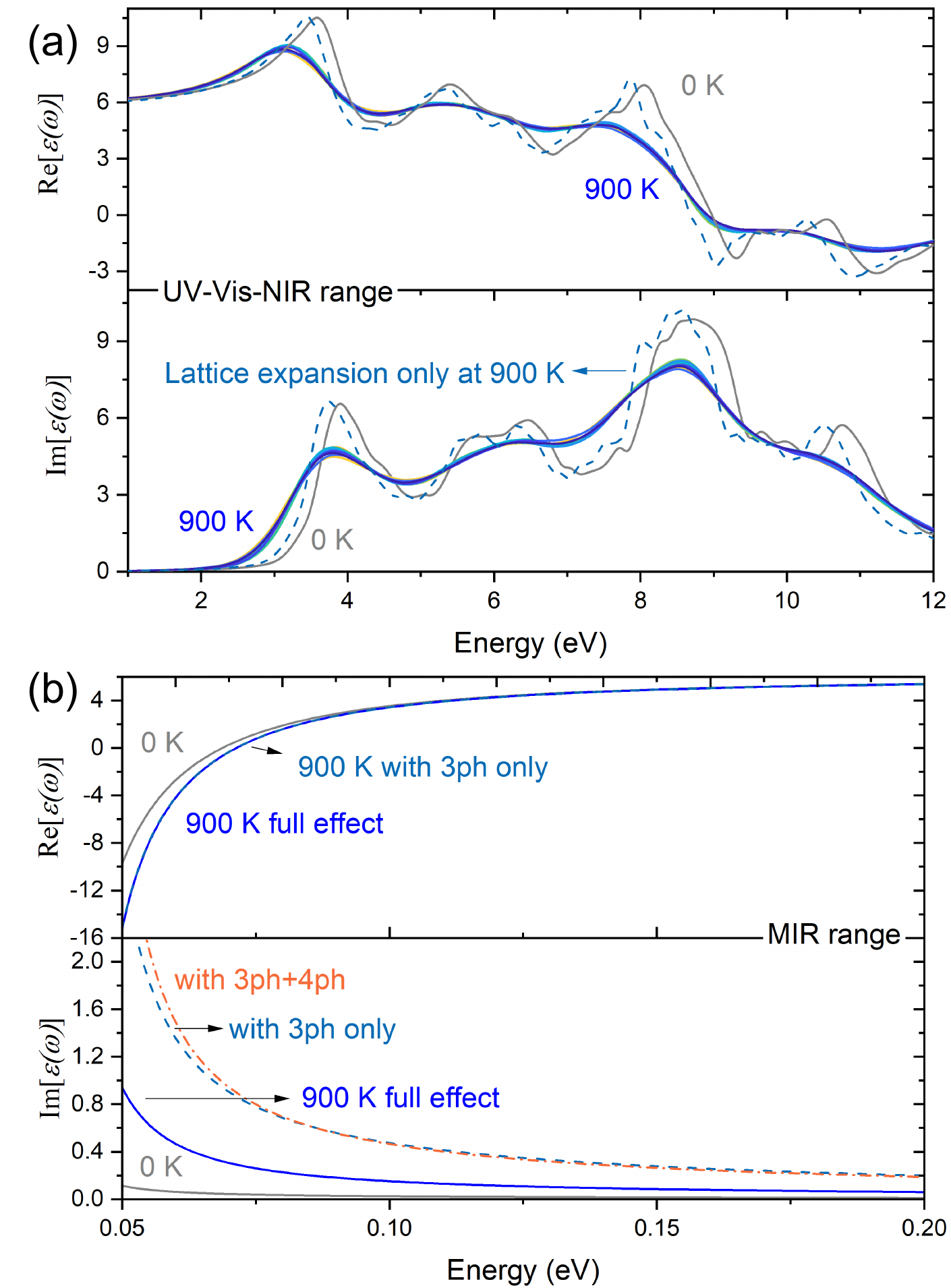}
    \caption{Dielectric function $\epsilon(\omega)$ at 0~K (grey lines) and 900~K (blue lines). (a) $\epsilon(\omega)$ from electronic transition in UV-Vis-NIR range. Each colored solid line represents the dielectric function for one snapshot structure in AIMD simulations. Dashed lines are $\epsilon(\omega)$ calculated on relaxed structure at 900~K. (b) $\epsilon(\omega)$ from phonon resonances in MIR range. Damping factor for 0~K is approximated by scattering rates at 5~K. Dashed cyan lines represent $\epsilon(\omega)$ with only 3ph scattering included. Dash-dot orange line represents $\epsilon(\omega)$ with 3ph+4ph scattering included. Solid lines represent $\epsilon(\omega)$ with both 3ph and 4ph scattering and phonon renormalization. In the upper panel of (b), dashed line nearly overlaps with solid line.}
    \label{epsilon}
\end{figure}

With the temperature-dependent energy carrier spectrum, we then proceed to calculate dielectric function in UV-Vis-NIR and MIR range:
\begin{enumerate}
    \item Dielectric function due to electronic transitions in UV-Vis-NIR range is the average of dielectric functions of all the collected snapshots~\cite{epsilonPRB2008}. Each snapshot is a perturbed supercell structure in AIMD simulation. We note that calculating dielectric function of a large supercell (192 atoms in our case) is nontrivial as the enlarged supercell structure has band folding in reduced Brillouin Zone. To address this issue and to reach convergence in DFT calculations, we include 2560 empty bands to allow sufficient electronic transitions in our first-principles calculations.
    \item For the dielectric function due to phonon resonances, we consider the recent theoretical advancements that single out the importance of phonon renormalization and higher-order anharmonicity in the prediction of optical phonon scatterings~\cite{Tong2020prb,han2022prl}, where optical phonon frequencies are renormalized (see Fig.~\ref{dispersion}(b)) and the phonon-phonon scattering has two-channel contributions: $\tau^{-1}=\tau^{-1}_{\rm 3ph}+\tau^{-1}_{\rm 4ph}$, i.e., three- ($\tau^{-1}_{\rm 3ph}$) and four-phonon ($\tau^{-1}_{\rm 4ph}$) scattering. The renormalized phonon energy and the effective third- and fourth-order force constants~\cite{TDEP2011,TDEP2013IFCs} are all obtained through AIMD simulations in this work, as to be consistent with our treatment on electronic band structure.
\end{enumerate}

\begin{figure*}[ht]
    \centering
    \includegraphics[width=6.8in]{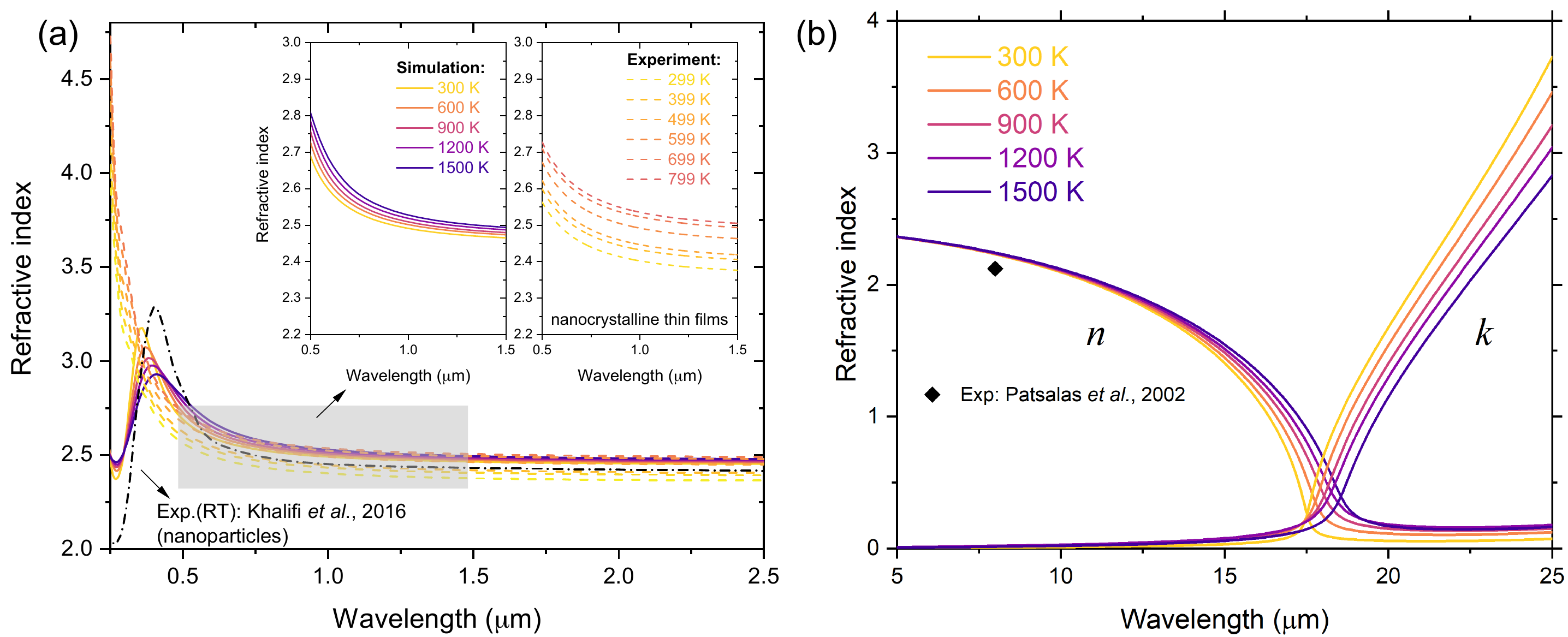}
    \caption{Temperature-dependent full spectrum refractive index. (a) Simulated refractive index (solid colored lines) with our ellipsometry measurement (dashed colored lines) up to 2.5~$\rm \mu m$ and from 300~K to 800~K, and an earlier measurement (black dash-dotted line) on CeO$_2$ nanoparticles~\cite{Khalifi2016} at room temperature. Imaginary part $\kappa$ is nearly zero in this range. (b) Simulated refractive index (solid colored lines) with an earlier measurement~\cite{Patsalas2002} (black diamond) in MIR range.}
    \label{spectrum}
\end{figure*}

Our calculated dielectric function $\epsilon(\omega)$ for the whole spectrum is presented in Fig.~\ref{epsilon}. In this plot, we compare the real and imaginary parts of the complex dielectric function at 0~K and 900~K. In UV-Vis-NIR range shown in Fig.~\ref{epsilon}(a), the first absorption peak in $\rm Im[\epsilon(\omega)]$ is correlated with the band gap reduction we present in Fig.~\ref{dispersion}(a): at higher temperature, lower photon energy is required to have interband transitions. This indicates that the material becomes more metallic. Another signature of temperature effect is the broadened peaks in both $\rm Re[\epsilon(\omega)]$ and $\rm Im[\epsilon(\omega)]$, which can be attributed to the structure disorder induced by stronger lattice vibrations. Note that this is captured in our AIMD simulations and averaging procedure. In contrast, the $\epsilon(\omega)$ with only lattice expansion considered (dashed lines) at 900~K shows shifts in peak energies but no broadening. This behavior reconciles with our observation from Fig.~\ref{dispersion}(a) that lattice thermal expansion has minor effect on renormalized band structure. Turning to MIR range shown in Fig.~\ref{epsilon}(b), we find that $\rm Re[\epsilon(\omega)]$ becomes more negative with temperature while $\rm Im[\epsilon(\omega)]$ decreases. This suggests that at higher temperature CeO$_2$ has larger dielectric loss with larger damping factor. Also, we observe that $\rm Im[\epsilon(\omega)]$ has larger temperature dependence while the difference in $\rm Re[\epsilon(\omega)]$ is marginal. At 900~K, using conventional approach that includes only three-phonon (3ph) scattering and no phonon renormalization effect (dashed cyan lines), $\rm Re[\epsilon(\omega)]$ is almost unaffected but $\rm Im[\epsilon(\omega)]$ is larger compared to the full effect. Further inclusion of four-phonon (4ph) scattering gives similar results (dash-dot orange line). Phonon renormalization then brings both 3ph and 4ph scattering back to a lower level (solid blue line). This is understood as phonon renormalization weakens the phonon scattering, and in the case of CeO$_2$ this effect is stronger than the sole inclusion of 4ph scattering to correct the conventional 3ph approach.

With the full spectrum temperature-dependent dielectric function $\epsilon(\omega)$, we can easily relate to refractive index, a more relevant optical quantities in applied fields, by using the relation: $\epsilon=(n+i\kappa)^2$ where $n$ and $\kappa$ are the real and imaginary part of the complex refractive index, respectively. To validate our first-principles computed results, we also measure $n$ and $\kappa$ at different temperatures using ellipsometry, where the measured amplitude ratio $\phi$ and phase difference $\Delta$ between the $p$- and $s$-polarizations are fitted into a Cauchy dispersion model to obtain the refractive index as a function of wavelength. The experiments are performed on a CeO$_2$ thin film sample that is deposited on a STO (001) substrate kept at 600~\textcelsius~by using pulsed laser deposition (PLD) with a KrF excimer laser. Details of this optical measurement and sample preparation are presented in Supplemental Materials~\cite{supply}. Our calculated results and experimental measurements are shown in Fig.~\ref{spectrum}. Two other literature measurements are also presented for comparison~\cite{Khalifi2016,Patsalas2002}. Very good agreement in temperature evolution is observed in the insets of Fig.~\ref{spectrum}(a). For wavelength longer than 0.5~$\mu \rm m$, refractive index $n$ increases with rising temperatures and this trend is consistent up to 800~K in experiment. This implies that the material is more reflective at higher temperatures. We note that the variation of $n$ is greater than what is predicted by first-principles calculations, though the numbers are in a reasonable range. Possible reasons include additional temperature-dependent effects that are not fully captured between our model and our CeO$_2$ nanocrystalline thin films. The first-principles calculation has assumed a perfect crystal structure. Another signature is the redshift of the first absorption peak around 0.4~$\mu \rm m$ or 3~eV, which is due to band gap reduction at higher temperatures. This change is consistent across the energy spectrum in Fig.~\ref{dispersion}(a), dielectric response in Fig.~\ref{epsilon}(a) and finally the optical response described in Fig.~\ref{spectrum}(a).

To summarize, we have established a first-principles framework for calculating temperature-dependent optical responses in full spectrum of semiconductors. This is enabled by a parallel treatment of electrons and phonons at finite temperature using AIMD. On top of thermally relaxed structure extracting from NPT ensemble, we capture the thermal perturbations from further NVT ensemble configurations and obtain renormalized electron and phonon energies. The dielectric function for the whole spectrum is then computed by electronic transitions in UV-Vis-NIR and Lorentz oscillator model in MIR range. The computed refractive index as a function of temperature is in good agreement with our own ellipsometer measurement on CeO$_2$ thin film. The first-principles methodology demonstrated in this study can have important implications in both optics and thermal radiation community.
\\
\\
\\

\begin{acknowledgments}
We acknowledge the support from the Defense Advanced Research Projects Agency under contract number HR00112190006. The views, opinions and/or findings expressed are those of the author and should not be interpreted as representing the official views or policies of the Department of Defense or the U.S. Government. 

X. R. and Z. H. thank Professor Thomas Beechem at Purdue University for helpful discussions on dielectric function of supercell structures. Simulations were performed at the Rosen Center for Advanced Computing (RCAC) of Purdue University.
\end{acknowledgments}


\bibliography{Reference}

\end{document}


\title{Supplemental Material for ``Temperature-Dependent Full Spectrum Optical Responses of  Semiconductors from First Principles"}

\author{Zherui Han}
 \affiliation{School of Mechanical Engineering and the Birck Nanotechnology Center,\\
Purdue University, West Lafayette, Indiana 47907-2088, USA}
\author{Changkyun Lee}
 \affiliation{School of Electrical and Computer Engineering and the Birck Nanotechnology Center,\\
Purdue University, West Lafayette, Indiana 47907-2088, USA}
\author{Jiawei Song}
 \affiliation{School of Materials Engineering,
Purdue University, West Lafayette, Indiana 47907-2088, USA}
\author{Haiyan Wang}
 \affiliation{School of Materials Engineering,
Purdue University, West Lafayette, Indiana 47907-2088, USA}
\author{Peter Bermel}
 \affiliation{School of Electrical and Computer Engineering and the Birck Nanotechnology Center,\\
Purdue University, West Lafayette, Indiana 47907-2088, USA}
\author{Xiulin Ruan}%
 \email{ruan@purdue.edu}
 \affiliation{School of Mechanical Engineering and the Birck Nanotechnology Center,\\
Purdue University, West Lafayette, Indiana 47907-2088, USA}

\date{\today}

\maketitle

\newpage
\tableofcontents

\section{Computational details}

In this section, we cover the computational details in our first-principles calculations.

All calculations are done using Density Functional Theory (DFT), Density Functional Perturbation Theory (DFPT) or \textit{ab initio} molecular dynamics (AIMD) as implemented in the VASP package~\cite{VASP1993}. For structural optimization, we use a Monkhorst $k$-grid of $12\times12\times12$ with 520~eV plane wave energy cutoff. Force convergence is within $10^{-6}$~eV/\r{A}. As described in the main text, we use the local density approximation with Hubbard parameter (LDA+U) approach by setting $U_{\rm eff}=10~\rm eV$ in this study to correct our band structure~\cite{HubbardU1991,LDA+UPRB2007}. Born effective charges are computed by DFPT and we get $\epsilon_{\infty}=5.927$, $Z^*_{Ce,xx}=5.502$ and $Z^*_{O,xx}=-2.751$ with the aid of Phonopy~\cite{phonopy}.

\begin{figure*}[h]
    \centering
    \includegraphics[width=5in]{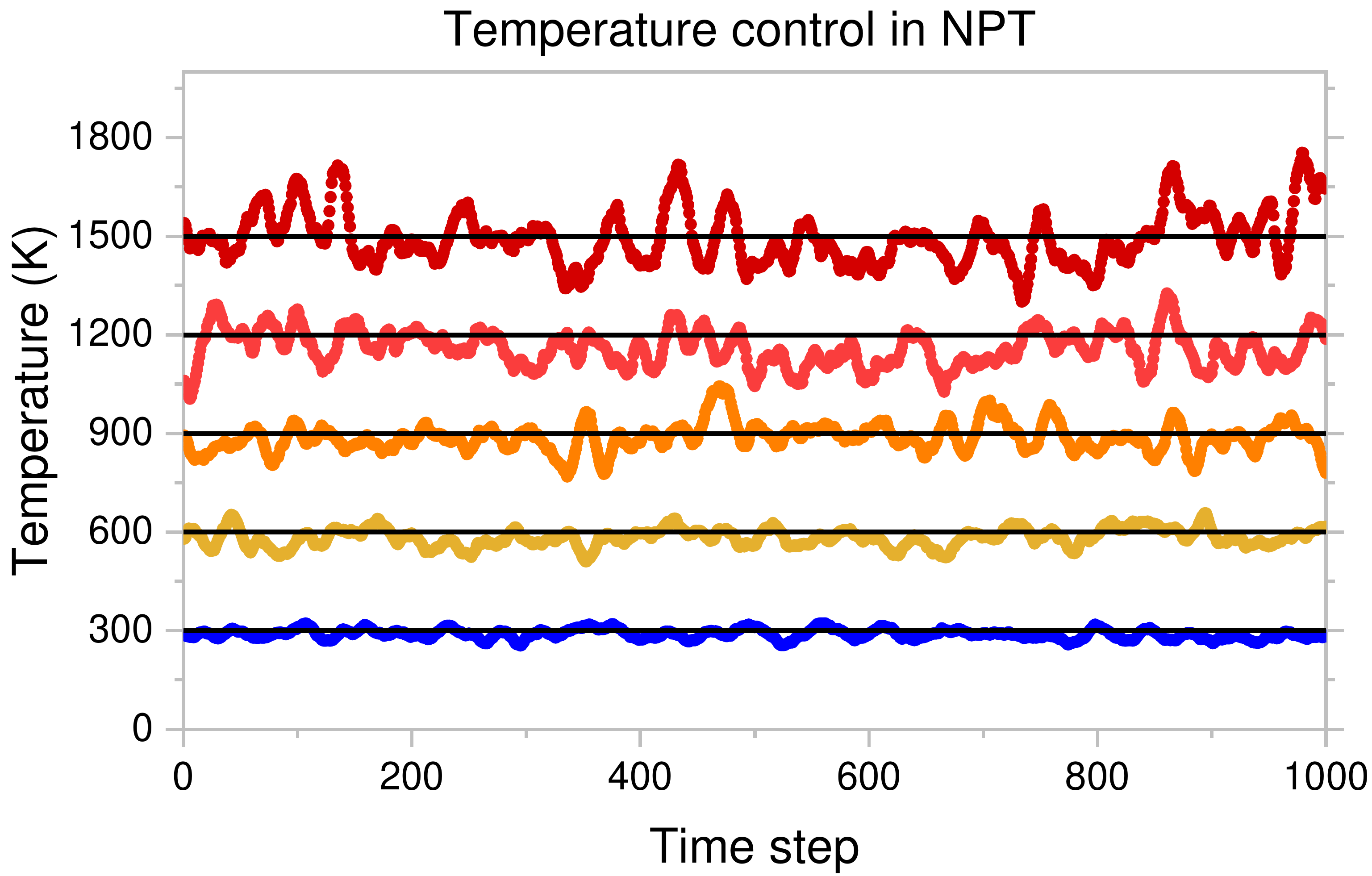}
    \caption{Temperature fluctuations of NPT ensemble. We report the productive run after reaching equilibrium and we use these steps to calculate the averaged lattice structure.}
    \label{NPT}
\end{figure*}

\textit{Ab initio} molecular dynamics are performed on a supercell structure consisting of 192 atoms constructed by $4\times4\times4$ CeO$_2$ primitive cells. For this size of supercell, only $\Gamma$ point is computed to accelerate the calculation. After reaching thermal equilibrium under NPT ensemble (zero external pressure) with Langevin thermostat, we use 1000 more steps to get averaged lattice structure at each temperature. An illustration of our temperature control in NPT ensemble is shown in Fig.~\ref{NPT}. Then, on relaxed structure we perform NVT ensemble simulations and after reaching equilibrium we use 2000 more steps to construct effective force constants~\cite{TDEP2013IFCs} at a time step of 2~fs. Harmonic force constants (HFCs) consider a cutoff radius of 6.31~\r{A}, the third-order force constants have 6~\r{A} and the fourth-order force constants have 4~\r{A} as cutoff.

Optical properties of $4\times4\times4$ CeO$_2$ supercell are computed in a DPFT manner. We use $\Gamma$-centered $k$-grid of $2\times2\times2$ and 2560 KS orbitals. This large number of empty bands for transitions is tested for convergence.

For phonon-phonon interactions in CeO$_2$, the Brillouin Zone (BZ) is discretized by $15 \times 15 \times 15$ $q$-mesh to evaluate three-phonon scattering rates using ShengBTE package~\cite{shengbte} and $10 \times 10 \times 10$ $q$-mesh four-phonon scattering rates using FourPhonon tool~\cite{han2021fourphonon}. In such process, we develop an in-house code to convert TDEP formats to formats that are compatible to our program including the treatment on long-range interactions and boundary conditions. Related tools will be made public when we update FourPhonon tool to next version.

\section{Ellipsometer measurement}

We performed the temperature dependent measurement using variable angular spectroscopic (VASE) integrated with heating cell (INSTEC, HCP621G). The cell is connected to the water cooler (INSTEC, mk2000) and digital heating controller to adjust targeting temperature manually. 

\section{Sample preparation}
\begin{figure*}[h]
    \centering
    \includegraphics[width=5in]{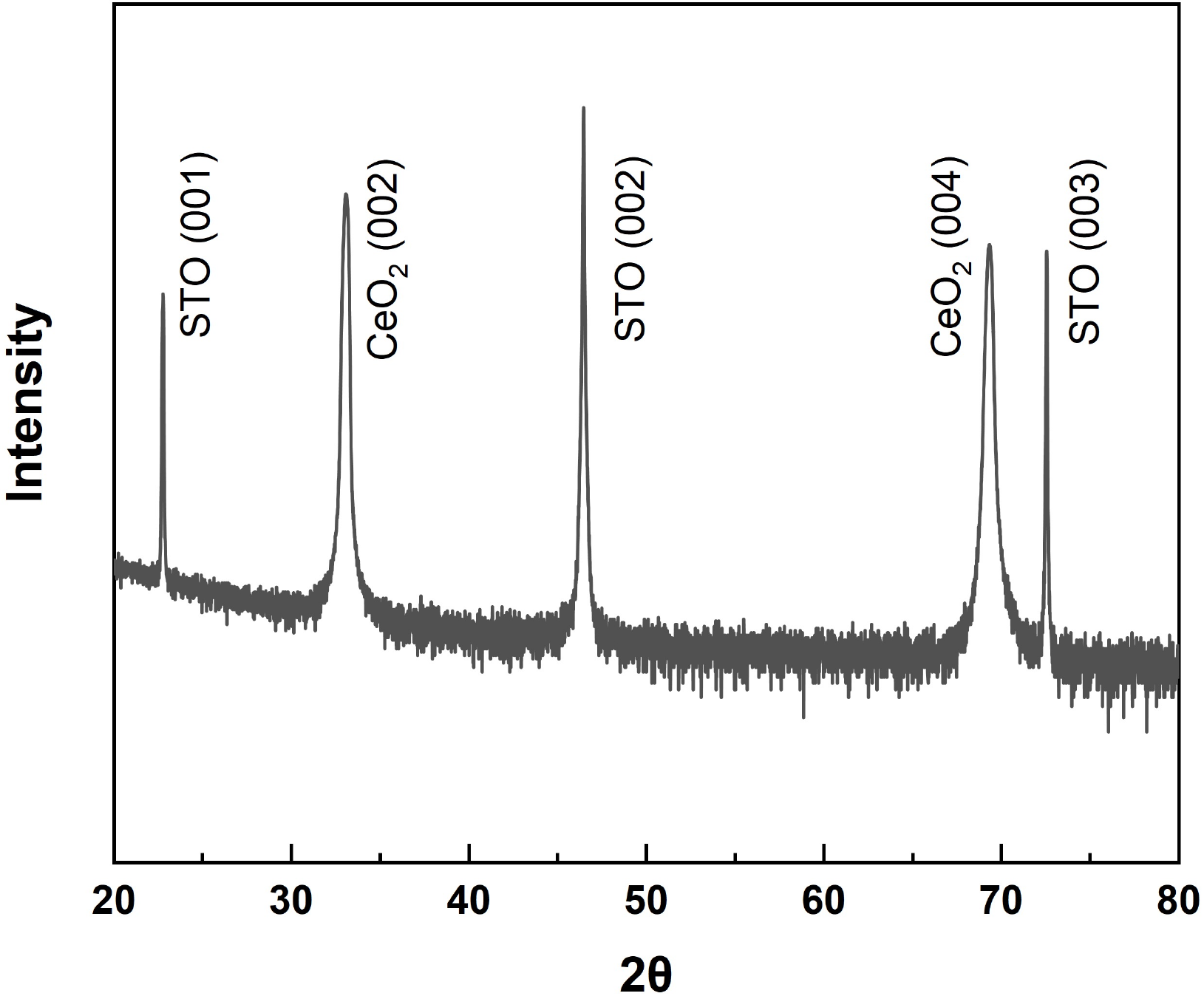}
    \caption{XRD of the CeO$_2$ film grown on STO (001) substrate.}
    \label{XRD}
\end{figure*}

The CeO$_2$ thin film was deposited on a STO (001) substrate by using pulsed laser deposition (PLD) with a KrF excimer laser (Lambda Physik, $\lambda$ = 248~nm). The substrate temperature was kept at $\SI{600}{\celsius} $ and a 20~mTorr oxygen pressure was used during deposition. After deposition, the temperature was naturally cooled down to room temperature at 20~mTorr oxygen partial pressure. The crystallinity and growth orientation was first characterized by X-ray diffraction (XRD). As shown in Fig.~\ref{XRD}, the CeO$_2$ film grows highly textured along (00l) direction on the STO substrate. The sharp peaks suggest good crystallinity of CeO$_2$ film, and no impurity peaks were observed in this CeO$_2$ film. 

\bibliography{Reference-si}